# Fitting and projecting HIV epidemics: Data, structure and parsimony

Brian G. Williams

South African Centre for Epidemiological Modelling and Analysis (SACEMA), Stellenbosch, South Africa
Correspondence to BrianGerardWilliams@gmail.com

**Abstract**

Understanding historical trends in the epidemic of HIV is important for assessing current and projecting future trends in prevalence, incidence and mortality and for evaluating the impact and cost-effectiveness of control measures. In generalized epidemics the available data are of variable quality among countries and limited mainly to trends in the prevalence of HIV among women attending ante-natal clinics. In concentrated epidemics one needs, at the very least, time trends in the prevalence of HIV among different risk groups, including intravenous drug users, men-who-have-sex-with-men, and commercial sex workers as well as the size of each group and the degree of overlap between them. Here we focus on the comparatively straight forward problems presented by generalized epidemics. We fit data from Kenya to a susceptible-infected model and then successively add structure to the model, drawing on our knowledge of the natural history of HIV, to explore the effect that different structural aspects of the model have on the fits and the projections.

Both heterogeneity in risk and changes in behaviour over time are important but easily confounded. Using a Weibull rather than exponential survival function for people infected with HIV, in the absence of treatment, makes a significant difference to the estimated trends in incidence and mortality and to the projected trends. Allowing for population growth has a small effect on the fits and the projections but is easy to include. Including details of the demography adds substantially to the complexity of the model, increases the run time by several orders of magnitude, but changes the fits and projections only slightly and to an extent that is less than the uncertainty inherent in the data. We make specific recommendations for the kind of model that would be suitable for understanding and managing HIV epidemics in east and southern Africa.

## Introduction

There are three rules of good modelling in epidemiology. First: stay as close to the data as possible. Second: include as much biology as possible. Third: keep it simple. Einstein expressed this more elegantly in observing that '… the supreme goal of all theory is to make the irreducible basic elements as simple and as few as possible without having to surrender the adequate representation of a single datum of experience'.[1]

Here we use HIV-prevalence data from ante-natal clinics (ANCs) in Kenya where the time trends and the quality of the data are typical of many African countries. We first fit a susceptible-infected (SI) model and explore ways of dealing with heterogeneity in the risk of infection and changes in overall risk behaviour. We then allow for the known Weibull survival distribution for people infected with HIV and for population growth. Finally we include details of the demography including the age dependent background mortality, the age-specific mortality for those infected with HIV but not on anti-retroviral therapy (ART), and the age-specific incidence of infection. We do not separate people out by sex and we do not include details of the age-matching of sexual partners.

## Data

While the ANC data for Kenya are of variable coverage and quality and the size of the epidemic differs significantly among countries in Africa, the data for Kenya are sufficient to establish a reasonable estimate of the national trend[2] and are representative of the trends seen in other countries in Africa.

For the compartmental models we only use data on the estimated time-trends in the prevalence of HIV infection in adults and the coverage of anti-retroviral therapy (ART). Because the prevalence data are from estimates of the underlying trend, fitted to smooth functions, we have added normally distributed random numbers to the estimates with standard deviations estimated from the published uncertainty bounds to simulate the actual data.

For the demographic model we need considerably more data. To determine the age-specific background mortality we use the current age-distribution of people in Kenya, the current population growth rate, and we assume a stable age distribution.[3] We use data on the age-specific incidence of infection from a study in South Africa[4] and data from the Cascade study[5] to estimate the survival of people as a function of the age at which they were infected.[6]

## Results

We first fit a basic SI model to the prevalence data from Kenya. To this we add additional structure to allow for heterogeneity in risk behaviour and for changes in behaviour over time. We then replace the implied exponential survival with a Weibull survival and then allow the population to grow over time. Finally introduce details of the demography.

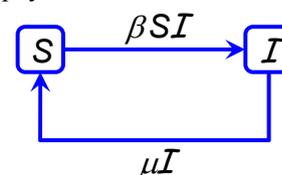

Figure 1. People are infected at a *per capita* rate $\beta I$, uninfected (susceptible) people die at a *per capita* rate $\delta$ and infected people die at a *per capita* rate $\mu$. If we feed $\delta S$ back into $S$ we can omit it and if we let $S + I = 1$, we are working with proportions.

**SI model**

The SI model is shown in Figure 1. Assuming that AIDS related mortality, $\mu$, is 0.1/year corresponding to a mean life expectancy of 10 years without treatment, the model has two parameters: the prevalence of HIV in 1970, when the model is started, and the transmission parameter, $\beta$, which determines the initial rate of increase. Fitting the *SI* model to the trend in the prevalence of HIV in Kenya gives the fit shown in Figure 2. As expected the model can fit the initial rise but not the peak and subsequent decline. Although the *SI* model only fits the early rise in infection it provides an estimate of 0.47/year for the initial growth rate and hence an estimate of 4.7 for $R_0 = \beta/\mu$.[7]

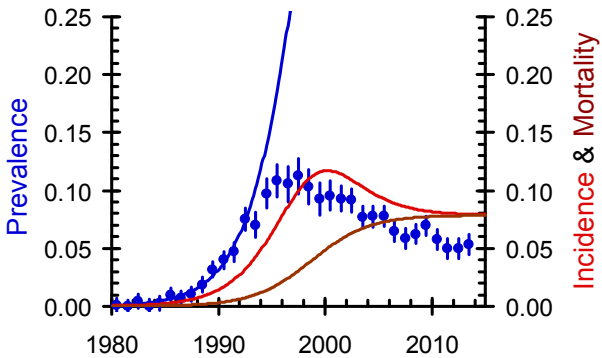

Figure 2. The SI model (Figure 1) fitted to the prevalence of HIV in Kenya.[2] The model is fitted to the data up to 1992. Blue: data with confidence limits; Green: prevalence; Red: annual incidence; Brown: annual mortality.

**Heterogeneity in risk**

In order to fit the peak prevalence of HIV we need to allow for heterogeneity in the risk of infection among people.[8] Given that high risk people are likely to be infected before low risk people, we assume that the transmission parameter, $\beta$, declines as the prevalence increases[9] and the rate at which this happens is varied to fit the data. If the functional form of this relationship is

$$\beta^* = \beta e^{-(\alpha P)^n} \qquad 1$$

then with $n = 1$ transmission declines exponentially with prevalence and as $n \to \infty$ the relationship approaches a step function. If we assume, instead, that a proportion $P^* = 1/\alpha$ are at risk while the rest are at no risk we have

$$\beta^* = \beta(1 - \alpha P) \qquad 2$$

(and $\beta^* = 0$ if $\alpha P > 1$) so that $\beta^*$ declines linearly from $\beta$ when $P = 0$ to $0$ when $P = 1/\alpha$. Replacing $\beta$ by $\beta^*$ in Figure 1 the model has two variable parameters: $\beta$ and $\alpha$.

Figure 3 shows the best fit to the data up to 1995 with $n = 1$, 2 and 4 using Equation 1 and using Equation 2. In all cases the prevalence curve lies precisely under the mortality curve because we are assuming exponential survival, and therefore a constant hazard of death, so that with an annual mortality rate of 10% for those infected with HIV the mortality is always 10% of the prevalence.

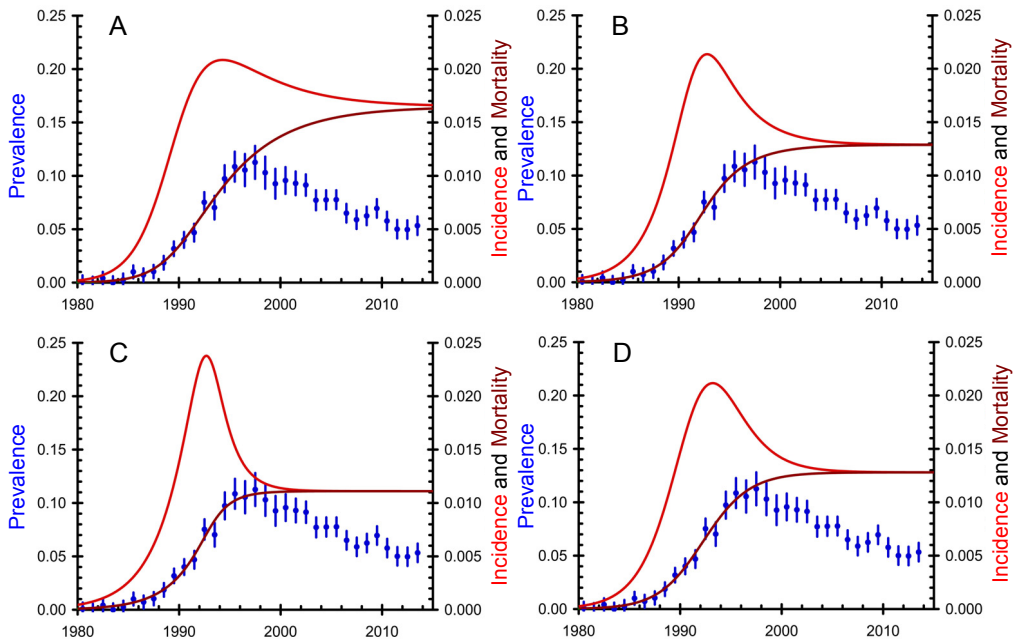

Figure 3. The SI model (Figure 1), allowing for heterogeneity in the risk of infection, fitted to the prevalence of HIV. A: $n = 1$; B: $n = 2$; C: $n = 4$ (Equation 1) and D: linear decline (Equation 2). Blue: data with confidence limits; Red: annual incidence; Brown: annual mortality. The prevalence curve (primary axis) lies exactly beneath the brown mortality curve (secondary axis); see text for details.

All four models (Equations 1 and 2) give equally good fits to the data up to 1995 so that the prevalence data alone are insufficient to decide among them. However, there are significant differences in the implied incidence curves and in the asymptotic prevalence. The peak value of the incidence increases from 2.1% p.a. when $n = 1$ (Figure 3A) to 2.4 % p.a. when $n = 4$ (Figure 3C) and the peak of the incidence cure is much flatter when $n = 1$ and much sharper when $n = 4$. There are also significant differences in the asymptotic prevalence which falls from 16.5% when $n = 1$ (Figure 3A) to 11.1% when $n = 4$ (Figure 3C). Furthermore, the estimated values of $R_0$, calculated as $\beta/\mu$, fall from 6.3 when $n = 1$ to 4.6 when $n = 4$. However, the fitted curves for $n = 2$ (Equation 1; Figure 3A) and for the linear decline with prevalence (Equation 2; Figure 3D) are very similar. To explain this we plot the reduction in transmission as a function of the prevalence in Figure 4. In all cases the asymptotic prevalence is reached when the transmission is reduced by a factor of about 5 but the long tail on the exponential curve ($n = 1$; Figure 4, blue line) gives a significantly higher asymptotic prevalence than the short tail on the Gaussian curve ($n = 2$; Figure 4, red line) but the latter is close to the asymptotic prevalence assuming that transmission declines linearly with prevalence (Figure 4, green line). In the models that follow we shall use Equation 1 with $n = 2$ to allow for heterogeneity in the risk of infection but note that the functional form of this relationship makes a significant difference and an epidemiological basis for choosing this function would significantly improve our confidence in the results.

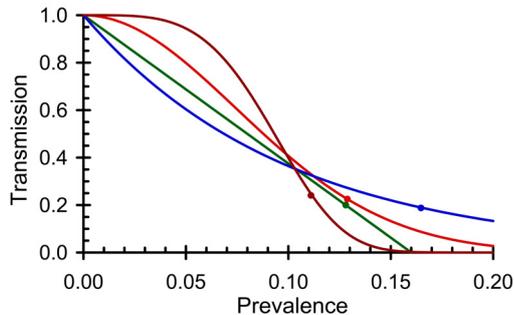

Figure 4. Transmission as a function of prevalence. Blue: $n = 1$; Red: $n = 2$; Brown: $n = 3$; Green: Linear. (See Equations 1 and 2). Dots indicate the reduction in transmission at the asymptotic prevalence.

**Control**

It is clear from Figure 3 that as the mortality rises the incidence, but not the prevalence, falls and there must have been a decline in the force of infection, after about 1995, which is not attributable to the natural history as captured by the model even allowing for heterogeneity in infection and, of course, mortality. This decline in risk over time could result from changes in people's behaviour as they become more aware of HIV or from changes bought about by external interventions resulting in increased condom use, delayed age of sexual debut, having fewer partners, changes in peoples movement and so on.

Whatever the reason for the decline in prevalence after 1997 we can add an external 'control' by letting the force of infection decline over time in a logistic fashion so that we replace $\beta^*$ in Figure 1 by $\beta^{\dagger}$ where

$$\beta^{\dagger}(t) = \beta^*(t) \left[ (1-\alpha_c) \frac{e^{\rho_c(t-\tau_c)}}{1+e^{\rho_c(t-\tau_c)}} + \alpha_c \right] \qquad 3$$

The term in square-brackets falls from 1 before the epidemic starts, converges to an asymptote $\alpha_C$, at a rate $\rho_C$, reaching half the asymptotic value at time $\tau_C$.

We do not specify the reason for this decline in the force of infection but the model will tell us by how much the force of infection must have changed to fit the data and the challenge is to find an explanation for a change of this magnitude. The model now has six variable parameters; $\mu$ and $\beta$ (Figure 1), $\alpha$ (Equations 1 or 2) plus $\alpha_C$, $\rho_C$ and $\tau_C$ (Equation 3). Fitting this model to the trend in the prevalence of HIV gives the result shown in Figure 5.

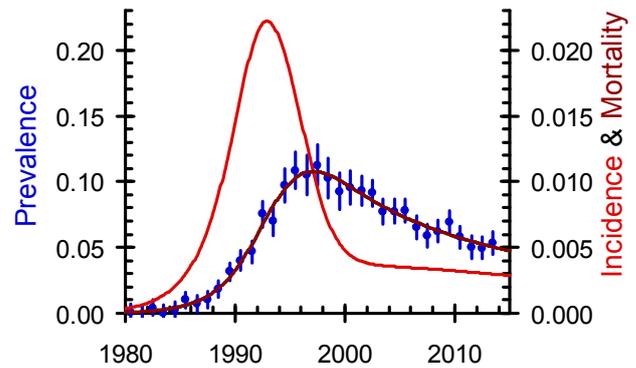

Figure 5. The model in Figure 1 allowing for heterogeneity in the risk of infection and for reductions in the risk of infection over time, fitted to the prevalence data. The prevalence curve (primary axis) lies exactly beneath the brown mortality curve (secondary axis); see text for details.

The fit to the data implies that the risk of infection fell by 85% in the four years between about 1991 and 2002 but has remained constant since then as shown in Figure 6.

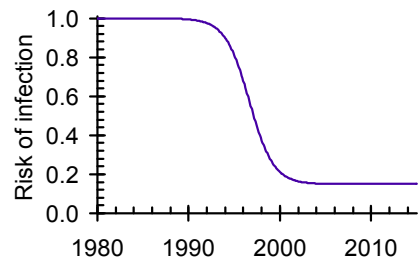

Figure 6. The decline in the risk of infection implied by the decrease in the prevalence of infection in Figure 5 from the model including control as defined in Equation 3.

**Confounding**

There is now a further problem that must be addressed. Heterogeneity in the risk of infection and changes in the risk of infection over time can both be used to control the peak prevalence in the model even though changes in the risk of infection over time although only some level of control can lead to a decline in incidence and prevalence. To illustrate this we go back to the model in Figure 1 keeping the control term (Equation 3) but leaving out the effect of heterogeneity in risk (Equation 1). This gives the fit shown in Figure 7 and the implied reduction in the risk of infection over time shown in Figure 8. The fits and implied incidence in Figure 5 and Figure 7 are similar but the effect of control shown starts about three years earlier, the control takes effect more slowly and the asymptotic risk reduction is 90% rather than 85% (compare Figure 6 and Figure 7).

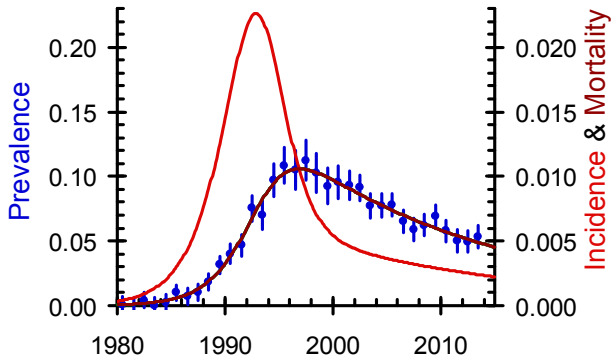

Figure 7. The model in Figure 1, allowing for reductions in the risk of infection over time but not heterogeneity in the risk of infection, fitted to the prevalence data. The prevalence curve (primary axis) lies exactly beneath the brown mortality curve (secondary axis); see text for details.

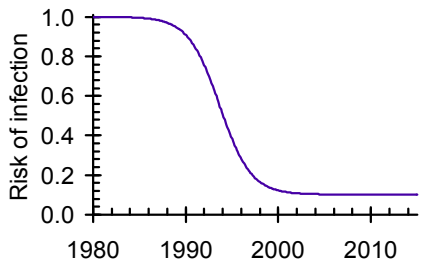

Figure 8. The decline in the force of infection implied by the decrease in the prevalence of infection in Figure 5 from the model including the control implied in Equation 3.

Both models fit the prevalence data in 2015 (5%) and then continue to fall. Allowing for heterogeneity and control the prevalence falls to 1.5% in 2050 but if we allow for control only the prevalence falls more quickly reaching 0.8% in 2050 (data not shown).

While the data, in themselves, do not allow us to decisively separate the effect of heterogeneity in risk and in control the long term consequences are significantly different and this must be considered when making forward projections.

These data remind us that, with this model, the initial rate of increase in the incidence is $\beta - \mu$ so that $\beta \approx 0.5$/year and $\mu \approx 0.1$/year the initial rate of increase in the prevalence of infection is about 0.4/year and the prevalence will double every $\ln(2)/0.4 = 1.7$ years. Even if transmission were stopped entirely the prevalence would fall at about 0.1 per year so that the incidence would halve every $\ln(2)/0.1$ years = 7 years. If we were to assume a value of $R_0 = 5$ and interventions that reduced transmission by 90% it would take about 90 years to reduce the prevalence by a factor of 100.

**Weibull survival**

Survival after infection with HIV is not exponential, which would imply a constant hazard, but rather follows a Weibull distribution with a shape parameter a little greater than 2 implying that mortality increases more or less linearly with time since infection. Detailed data on the survival with HIV, but without ART, as a function of the age at infection are available from the CASCADE cohort.[5] Fitting these data to Weibull survival functions for the probability of surviving for *t* years after being infected

$$W(t \mid \mu, \sigma) = 2^{-(t/\mu)^\sigma} \qquad 4$$

gives the median survival, $\mu$, and shape parameter, $\sigma$, as

$$\mu = 16.492 - 0.172a \qquad 5$$

$$\sigma = 2.673 - 0.011a \qquad 6$$

where *a* is the age at infection in years.[6] In most generalized epidemics the incidence of infection peaks between the ages of about 25 and 30 years, earlier in women than in men. For a Weibull distribution the mean survival *m* is

$$m = \mu \Gamma\left(1 + \frac{1}{\sigma}\right) \Big/ \ln(2)^{1/\sigma} \qquad 7$$

so that for those infected at 25 years of age the mean survival is 10.6 years while for those infected at 30 years of age it is 9.8 years. For the purposes of this study we assume a mean survival of 10 years and a shape parameter $\sigma$ of 2.3. The survival curve for this distribution is close to a $\Gamma$-distribution, with a mean survival of 10 years and a shape parameter of close to 4. This in turn means that a good approximation to the survival distribution is obtained by using four compartments for those infected with HIV and a median duration in each compartment of 2.5 years or a rate of progression from one compartment to the next of 1/2.5 or 0.4 per year as shown in Figure 9.

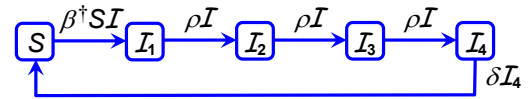

Figure 9. SI model with the addition of four stages of HIV infection to allow for the survival distribution of people infected with HIV.

The model in Figure 9 gives the fit to the data shown in Figure 10. The peak incidence is a little lower at about 1.8% *p.a.* (Figure 10) compared to 2.2% *p.a.* (Figure 5). The current (2015) incidence is higher at 0.31% *p.a.* (Figure 10) compared to 0.23% *p.a.* (Figure 5).*

Figure 11 shows the implied reduction in the force of infection using the model in Figure 9 and the fit in Figure 10. The reduction in the force of infection is now 85% close to the value of 83% implied by the model in Figure 5 although it starts about two years earlier.

The most important change resulting from the inclusion of four stages of HIV-infection (Figure 9) is that it introduces a delay of about 10 years between the rise and fall of the incidence curve and the rise and fall of the mortality curve as is immediately clear by comparing Figure 5 and Figure 10. This is important if we wish to model the initial trends, the impact of interventions, and future projections accurately.

---

* The peak prevalence, by definition corresponds to the point at which mortality exceeds incidence (see Figure 5, Figure 7 and Figure 10). While prevalence peaks when mortality exceeds incidence this only implies convergence to a lower asymptote, it does not imply elimination.

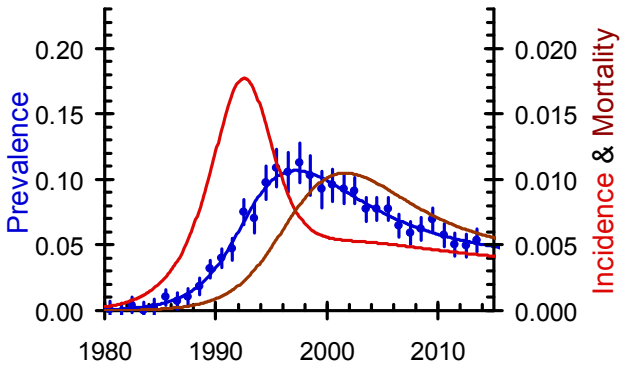

Figure 10. The model in Figure 9 fitted to the prevalence data including four stages of HIV infection to allow for the Weibull survival distribution for people infected with HIV.

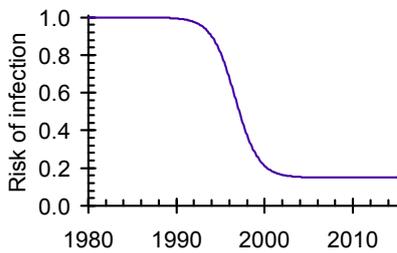

Figure 11. The decline in transmission implied by the decrease in the prevalence of infection in Figure 10 from the model in Figure 9.

**Population growth**

So far all the models have expressed the data in terms of the number of people in each state as a proportion of the total population. In practice the total population is increasing in most countries badly affected by the epidemic of HIV and we need to know how this affects the model fits and predictions. We therefore extend the model in Figure 9 to allow the population to grow by separating the background mortality, the HIV progression and AIDS deaths, and the birth rate, as shown in Figure 12.

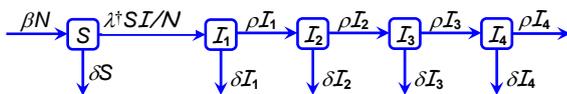

Figure 12. The model shown in Figure 9 but separating births ($\beta$), background mortality ($\delta$) and HIV progression and death ($\rho$).

We now use the model shown in Figure 12, set the crude birth rate to 3.6% per year and the background mortality to 1.2% per year so that the population grows at a rate of 2.4% per year in the absence of HIV. The gives the fit shown in Figure 13.

Allowing the population to grow introduces further changes. Comparing Figure 13 with Figure 10 we see that allowing for population growth increases the incidence, as a proportion of the total population, by about 10% and decreases the mortality, as a proportion of the total population by about 10% because the incidence rises when the population is smaller and the mortality rises when the population is greater.†

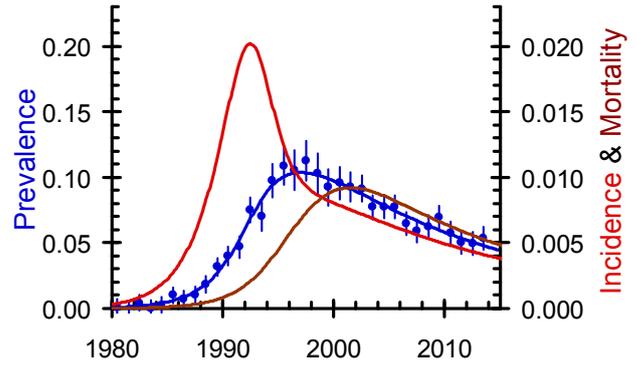

Figure 13. The model in Figure 12 making allowance for the growth of the population and plotting the rates as a proportion of the total population at a given time. Blue: data with confidence limits; Green: fitted prevalence. Red: annual incidence; Brown: annual mortality.

**Treatment**

To assess the impact of treatment we need to consider the mortality of people on ART. With modern, triple-therapy the survival of people on ART, is close to the survival of HIV-negative people even for those that start late in the course of their HIV infection.[10] In 30 year old adults in Kenya the mortality rate, in the absence of HIV, is about 1.2% per year.[3] In a study with a mean duration of follow up of about 3 years based on data from Uganda, Malawi and Kenya,[11] the annual AIDS related mortality in people infected with HIV was 0.4%. 0.6%, 0.9%, 1.9% and 7.4% in those starting ART at $CD4^+$ cell counts > 500, 350–500, 200–350, 50–200 cells/μL. Only in the two lowest $CD4^+$ cell count categories is mortality higher than the average mortality in HIV-negative people. For the model including treatment we will assume that those on ART experience the same mortality as those not on ART.

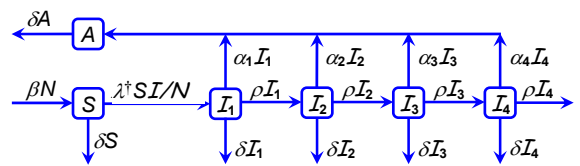

Figure 14. The model shown in Figure 12 but allowing people to start ART at different rates in each HIV-compartment and assuming that survival on ART is the same as in HIV-negative people.

In order to fit the data to the model in Figure 14 we vary the previous six parameters needed for the model without ART: the prevalence of HIV at the start of the epidemic, two parameters defined in Equation 1 and the three parameters defined in Equation 3. We also let the coverage of ART increase logistically so that the proportion of the whole population that is on ART is given by

$$A(t) = \alpha_a \frac{e^{\rho_a(t-\tau_a)}}{1+e^{\rho_a(t-\tau_a)}} \qquad 8$$

where the coverage converges to an asymptote of $\alpha_a$, at a rate $\rho_a$ reaching half the asymptotic value at time $\tau_a$. We also have to decide on the coverage in each stage of infection. Ideally one would use field data to inform these numbers but for the present purposes we first assume an overall coverage rate of 90% but only for people in Stage 4 as indicated in Figure 14 so that $\alpha_4 = 0.90$. This gives a reasonable but not quite adequate fit so we allow for coverage of people in Stage 3. This gives a good fit with $\alpha_3 = 0.63$ as shown in Figure 15.

---

† Note that an effect of population growth is that the peak prevalence now occurs 2.5 years before the mortality exceeds the incidence (see Figure 13).

Allowing for the reported coverage of ART we see that with about half of all HIV-positive people in Kenya on ART, the incidence has fallen by about 50% and the mortality by about 90%. One could, of course, achieve an equally good fit to the coverage data by assuming that more people are started in Stages 1 and 2 and fewer people in Stages 3 and 4; the important point is that the model allows us to explore the impact of starting ART earlier or later.

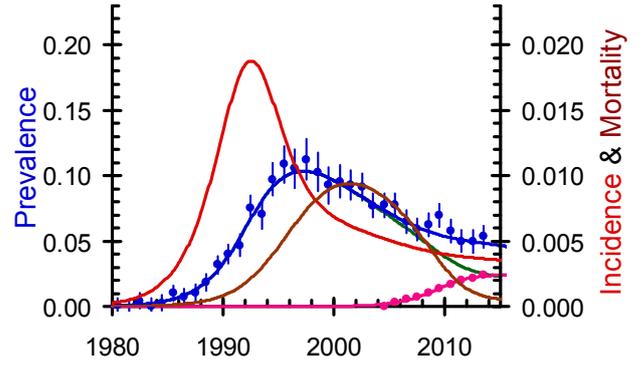

Figure 15. The model in Figure 14, allowing for the increasing coverage of ART but assuming that most of people start ART in Stage 4 and some in Stage 3. Blue: data with confidence limits; Green: fitted prevalence. Red: annual incidence; Brown: annual mortality; Pink: ART coverage.

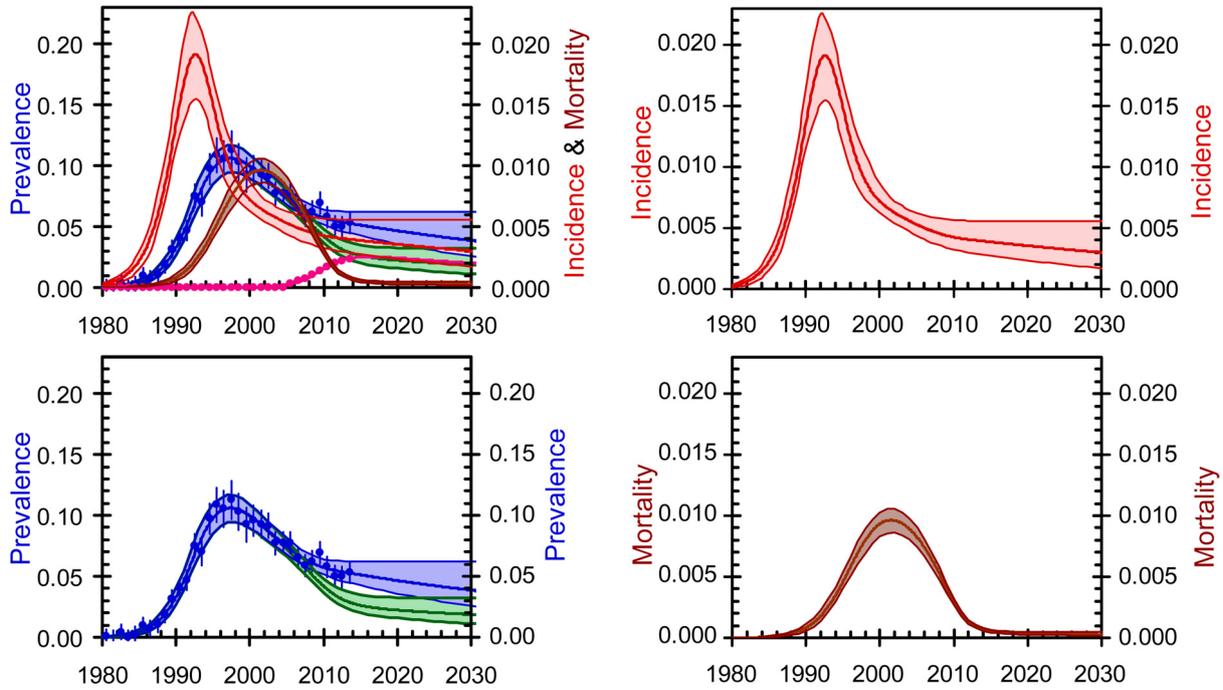

Figure 16. The fits in Figure 15 with 95% confidence bands on the fitted curves. A: All fitted curves; B: Annual incidence; C: Total prevalence, blue: Prevalence not on ART, green; D. Annual AIDS related mortality.

**Uncertainty**

Because the compartmental model is still relatively simple we can easily calculate uncertainties in the fitted parameters and fitted curves using a Markov-Chain Monte Carlo algorithm with a Gibbs sampler and binomial errors. This gives the results show in Figure 16. Even with relatively good data on the trends in the prevalence of HIV there is considerable uncertainty in the projected values. In 2030 the projected annual incidence is 0.30% (0.17% to 0.56%), the prevalence among those not ART, assuming that the coverage of ART remains at about its present level, is 1.8% (1.1% to 3.2%) and the annual mortality is 0.02% to 0.05%).

**Demography**

A full demographic model adds considerable complexity and it is important to investigate the extent to which this changes the fits and projections of the model. We need to allow for:
1. The age-specific mortality in people who are not infected with HIV, which may change over time.
2. The birth rate which may change over time and depends on the number of women of reproductive age.
3. The age-specific risk of infection.
4. The age-specific survival of people on ART.
5. The age distribution of people's sexual partners.

6. The difference in the risk of infection in men and in women.

Dealing precisely with the age-mixing of sexual partners is complicated and we will ignore this aspect of the demography for the present work. This gives the model illustrated in Figure 17.

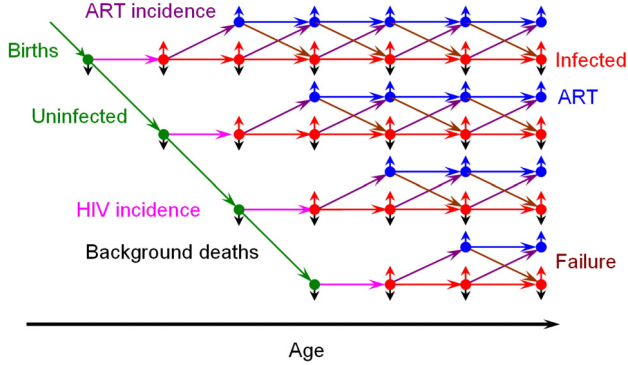

Figure 17. Overview of the demographic model. Green arrows: births and ageing of uninfected people; Pink arrows: incident infections; Red arrows: progression of people infected with HIV; Purple arrows: people starting ART; Brown arrows: people failing ART; Black vertical arrows: background mortality; Red vertical arrows: AIDS related mortality; Blue vertical arrows: mortality on ART. Each column of points represents people whose age is one time step greater than the previous column of points.

For the model in Figure 17 we need to use a time step of 0.1 year because the important rate processes are of the order of 1/year and models without demography show that this time step is sufficiently small to ensure stability of the solutions. We also use an age interval of 0.1 year. If we consider people of ages 0 to 100 years then the model in Figure 16 has $10^6$ states and that need to be updated from say 1970 to 2030 or 60 years and 600 time intervals. The number of states that need to be calculated over the course of one simulation is therefore of the order of $6 \times 10^8$. This then has to be repeated a sufficient number of times to ensure that the variable parameters provide the best fit to the observed data. A fuller discussion of the computational details of the demographic model is given in Appendix 1.

*Age-dependent mortality*

The age distribution will change over time but we assume here that the current age distribution is stable. We therefore take the current age distribution and the overall population growth rate, $r$, to estimate the age dependent mortality, $m(a)$,

$$m(a) = -\ln\left(\frac{a+1}{a-1}\right) - r \qquad 9$$

as shown in Figure 18.

Using Equation 9 implies negative mortality rates in those under the age of 15 years but this is a reflection of the fact that fertility has fallen over the last fifty years and the age-distribution is changing over time. Here we set the mortality to zero below the age of 15 years and then fit the mortality above the age of 15 years to an exponential so that the fitted mortality $m'(a)$

$$m'(a) = \alpha\left(1 - e^{-\beta(a-\gamma)}\right) \qquad 10$$

where the asymptotic value $\alpha = 0.047$/year, the rate of increase $\beta = -0.025$/year and the offset $\gamma$ is 14.7 years. Using Equation 10 and the population growth rate we back-calculate the stable age-distribution as a check and compare it to the current age-distribution (Figure 19) showing that the assumed and modelled age distributions are similar.

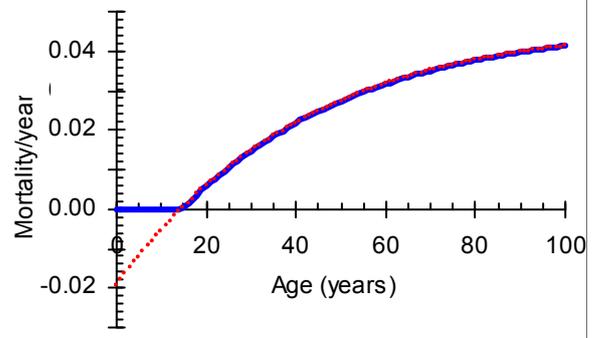

Figure 18. Age-specific mortality. Red dots: calculated using Equation 9; Blue line: fitted to data for those above the age of 15 years.

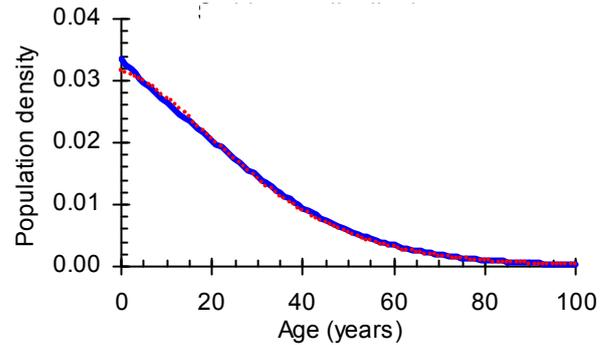

Figure 19. Red dots: current age distribution. Blue line: assumed stable age distribution.

As a further check we calculate the probability of surviving to any given age (Figure 20) from the full demographic model but without HIV. Because we have assumed that there is no mortality before the age of 15 years the curve is flat up to that age. Here we are interested in HIV in adults and this assumption will not affect the results significantly.

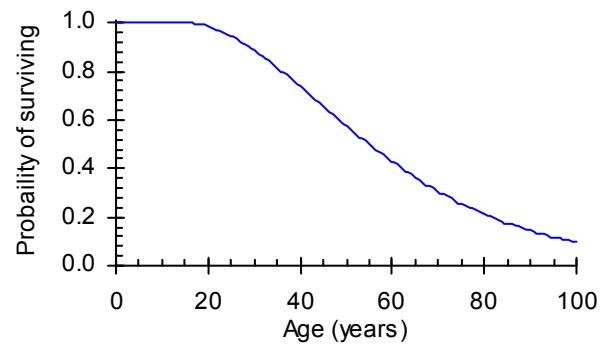

Figure 20. Probability of surviving to any given age in the absence of HIV.

As a final check we calculate the survival distribution of a person aged 30 years (Figure 21) and this agrees with the assumed survival based on the Cascade data[5] (Equations 5 and 6)

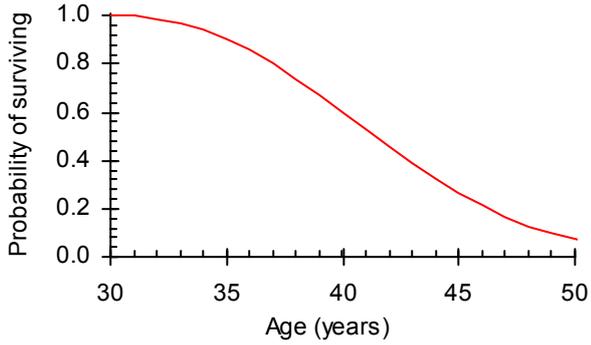

Figure 21. Calculated survival for those infected with HIV at 30 years of age.

The assumed age-specific incidence of infection (Figure 22)[4] implies that people become sexually active at the age of about 15 years, the risk of infection peaks at the age of about 32 years and falls to 10% of the peak value at the age of 60 years.

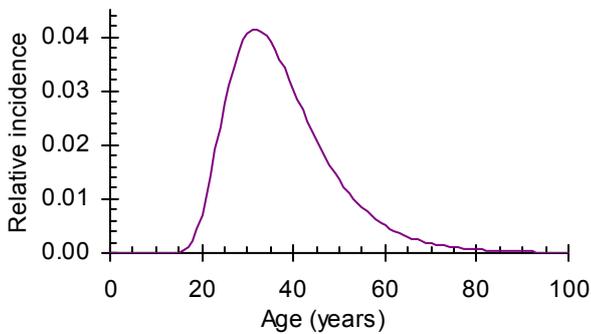

Figure 22. Assumed age-specific incidence of infection normalized to 1.

As a final check we run the model without HIV to confirm that it gives the right rate of growth of the population (Figure 23). The slope of the curve (2.4% *p.a.*) is equal to the assumed crude birth rate (3.6% *p.a.*) minus the assumed crude mortality rate (1.2% *p.a.*)

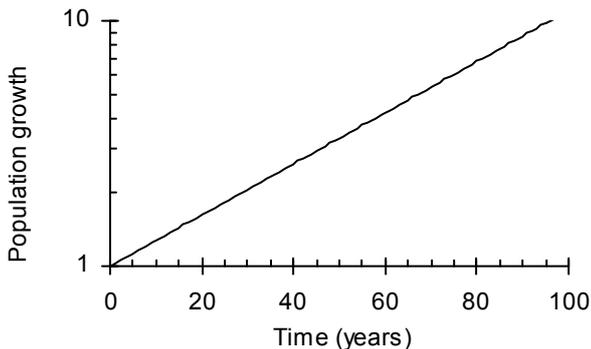

Figure 23. The growth of the population in the absence of HIV calculated from the demographic model. The slope of the curve is 2.4% per year.

We then use the demographic model to get the best fit to the prevalence data without ART. As before, we vary the six key parameters: the prevalence in 1970, the force of infection, and the heterogeneity, as well as the three parameters that define the fall in the force of infection over time (Equation 3). The fitted and implied trends for the best fit model are given in Figure 24.

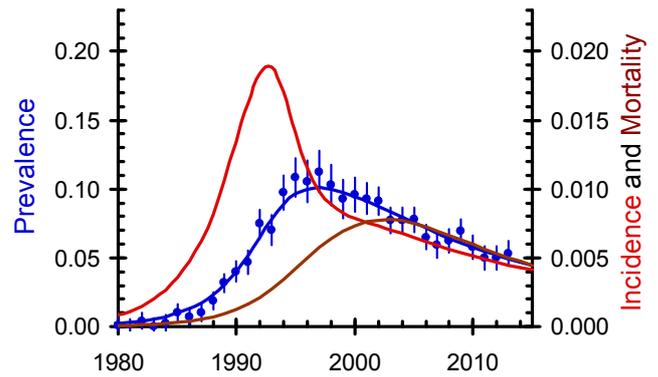

Figure 24. The best fit of the full demographic model, without ART.

The full demographic model gives a peak incidence of 1.9% *p.a.*, close to the value of 2.0% *p.a.* in the compartmental model, and about the same incidence of 0.4% in 2015. The full demographic model gives a peak mortality of about 0.8% *p.a.* compared to about 0.9%, and a current value of the mortality of 0.4% as compared to 0.5%. We can also compare the implied reduction in transmission over time for the compartmental model and the full demographic model. The compartmental model implies the transmission must have fallen by about 80% while the full demographic model implies that it fell by 74%.

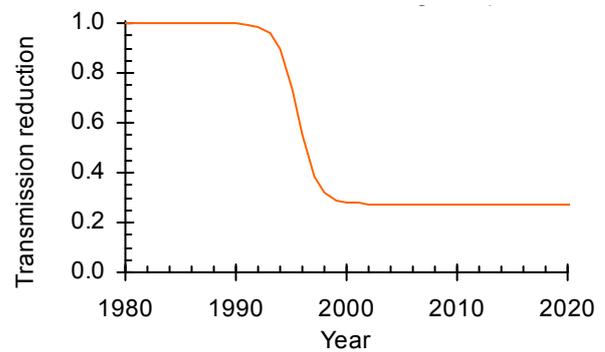

Figure 25. The implied decline in transmission for the full demographic model.

## Discussion

In this paper we start from the basic *SI* model and then add structure step-by-step. We are interested in knowing what effect different structural additions have, how important they are, which need to be included and to what extent they add to our understanding of the underlying biology.

The basic *SI* model describes the initial exponential rise in the prevalence but cannot fit the peak prevalence. It is therefore essential to allow for a decline in the average risk of infection as prevalence increases. The difficulty here is that a wide range of risk functions from an exponential to a step-function dependence on prevalence give equally good fits to the prevalence data but imply significantly different incidence and mortality rates. If data were available on the incidence or mortality it would help to resolve this problem but resolve it we must.

In some countries such as Kenya, illustrated here, and Zimbabwe[12] prevalence has fallen substantially and quite rapidly from a peak while in others, such as South Africa, the prevalence shows no sign of declining.[13] Where such a decline has been observed the reasons for the decline are unclear. It could be that people have changed their behaviour, and have sustained this change over time, as a result of a general awareness of the epidemic. It could be that people have changed their behaviour as a direct response to AIDS related deaths but an analysis of data from Zimbabwe suggests that this is not the case. Finally, it could be a result of the intrinsic dynamics of the transmission network but there are no convincing models that support this. We therefore allow for a logistic reduction in the transmission parameter. This allows us to fit the data and gives a measure of the extent to which transmission must have fallen. While it does not provide the answer to the question is defines the question more precisely.

The basic model, with only one compartment for infected people, implies exponential survival for those on ART which we know not to be the case. However, if one includes four stages of infection it is then possible to get a fairly accurate representation of the average survival of a person after they are infected with HIV. This has a significant effect on the implied incidence and introduces a delay of about 5 years between the rise and fall in the prevalence and the rise and fall in the mortality.

Allowing the population to grow, rather than working with proportions, increases the incidence and decreases the mortality, each by about 10%. While this change is small it is easily incorporated.

Where treatment has been made available the effect of this must be included and this of course will have a substantial impact on the time course of the epidemic. The problem is that few direct data are available on the actual coverage of ART in any country in the world. Since, in the short term, say five to ten years, the effect of providing ART is to leave the total number of people living with HIV more or less unchanged while the proportion of people not on ART does of course drop, the prevalence data alone are not sufficient to confirm the extent to which ART has been provided.

The full demographic model is much more computationally intensive increasing the number of states that have to be calculated for each run of the model increasing from about 360 in Figure 14 to about 600,000 in Figure 17 with a corresponding increase in the time taken to run the model and fit it to the data. However the fitted prevalence does not change and the implied mortality and incidence only change by about 10%.

## Conclusion

Simple, compartmental models fitted to time trends in the prevalence of HIV provide reliable estimates of the corresponding time trends in incidence and mortality from which projections can be made of the likely impact and cost effectiveness of interventions.

The models must first allow for heterogeneity in the risk of infection among people and second include sufficient structure to reflect the survival distribution for people with HIV but without ART and dealing with this appropriately remains the most important are of uncertainty.[8] Allowing the population to grow over time makes a small difference but does not increase the computational complexity of the model.

Including details of the demography such as the age-dependant mortality for those not on ART, age specific incidence of infection, and variation of survival on ART as a function of the age at infection change the implied incidence and mortality only slightly, and probably by less than the uncertainty in the data to which the model is fitted, but increase the computational burden by four orders of magnitude so that it is not worth using a full demographic model except in that happy situation where one has detailed data on the prevalence of infection as a function of age and, better still, reliable data on time trends in the incidence and mortality. With a simple compartmental model it is straight forward to estimate the uncertainty in the fits using a Markov Chain Monte Carlo approach while this would be even more demanding of computing power with a full demographic model.

In summary, simple compartmental models are sufficient to model generalized epidemics but a key unresolved issue is the nature and effect of heterogeneity in sexual behaviour and in sexual mixing patterns. Demographic models add substantially to the computational burden without changing the fitted and implied epidemiological trends significantly.

## Recommendations

Countries in east and southern Africa, where HIV is mainly spread through heterosexual contacts, need to be able to fit a suitable model to their data and use this to estimate current and predict future trends in the prevalence, incidence, mortality and ART coverage while including estimates of uncertainty. Based on this study we make the following recommendations.

**Use a compartment model**

Including details of the underlying demography changes the fits and projections by considerably less than the uncertainty that is inherent in the data while increasing the run-time for fitting the models from fractions of a second to many hours.

**Allow for heterogeneity in risk**

Include a term that allows the risk of infection to fall as the prevalence rises. A Gaussian relationship is recommended but a biological justification for a different functional form would be worth pursuing.

**Allow for changes in risk over time**

In some countries, but not all, it is clear that there has been a substantial decline in the risk of infection over time. A logistic decline in transmission gives a good fit to the data. However, the reasons for this decline or, in some places, for the absence of a decline demand further investigation.

**Allow for the known survival of people not on ART**

A four-compartment model for those infected with HIV gives a sufficiently accurate reflection of the average survival of those infected with HIV.

**Allow for treatment failure**

Although we have not included it here, allowing for four stages for people on ART, so that they move from infected stage $I_i$ to ART stage $A_i$ would allow one to vary the survival according to the stage of infection when treatment was started. If people fail treatment in stage $I_i$ they could then be returned to stage $I_i$.

**Consider including second-line treatment**

This could be included either explicitly, through a further set of four compartments, or implicitly by assuming that a proportion of people are on second-line treatment. Since there is clear evidence that with high coverage and good compliance, triple-therapy will eliminate drug resistance where it does arise,[14] this should be of secondary importance.

**Allow for population growth**

Although the effect of population growth is slight it can easily be included and is needed if we are to make estimates of the number infected and the number of incident cases and deaths.

**Allowing for age- and gender-specific rates**

Given the lack of reliable trend data in age-specific rates of infection of sexual mixing patterns it should be sufficient to distribute the total number of cases according to age and gender using measured distributions of prevalence, incidence and mortality

**Include ART coverage**

The ART coverage must be included; where possible as a function of disease progression.

**Uncertainty bounds**

These must be included as the uncertainty is likely to be significant.

**Outputs**

The model outputs should include current and projected trends in prevalence, on and off ART, incidence, the rate at which people start ART, and mortality, all with uncertainties. Including the prevalence off ART for the different stages of infection could be useful as the cost of treating AIDS related conditions will depend on the stage of infection.

# Caveat

ART has not been included in the full demographic model. While this adds further complexity and it might be interesting to explore this, it seems that the full demographic model is not generally needed and we have not pursued this further here.

# Appendix 1. Computational details of the demographic model

An overview of the demographic model is given in Figure 17. The detailed assumptions are as follows.

**Demography without HIV**

The age-specific death rate determines the mortality at each black arrow and these background mortalities, $\mathbf{B}_i$ at age $i$, also apply to the infected stages. These and the birth rate, $\beta$, remain fixed.

**Age specific incidence**

An estimate of the relative age-specific incidence is used to determine the age-specific transmission paramter $\mathbf{F}_i$, which we multiply by a factor $\lambda$ which will be varied to optimise the objective function.

**Age specific mortality with HIV**

The age-specific, AIDS-related mortality, $\mathbf{A}_i^j$, which depends on the age at infection $i$ and the current age $j$ so that the time since infection is $j - i$. This is chosen so that survival after infection follows a Weibull distribution as a function of the age at infection and the time since infection with the parameter values given in Equations 5 and 6.

**Age specific mortality on ART**

We let this be the same as the mortality in HIV-negative people.

**Rate of starting ART**

We start by keeping the rate of starting ART fixed but we could let it vary with age and/or with the time since infection and set it to $\mathbf{T}_i^j$. We have a separate parameter, $\sigma$, that scales the rate at which ART is rolled out so that we can decide when to introduce ART and to what coverage.

**Rate of failing ART**

We start by keeping the rate of failing (stopping) ART fixed but we let it vary with age and with the time since infection and we set it to $\mathbf{S}_i^j$. We have a separate parameter, $\phi$, that scales the failure rate but have not explored this here.

**Coded structure**

We assume that people can live for up to 100 years. With a time step of $dt$ we will need a $100/dt \times 100/dt$ population matrix and we will let the successive diagonal elements correspond to the ages of the uninfected people. $\mathbf{N}_i^i(t)$ is the number of susceptible people of age $idt$ at time $t$.

The elements above the diagonal correspond to people who are infected with HIV so that the element $\mathbf{N}_i^j(t)$, $j > i$, gives the number of people who were infected with HIV at age $i$ and are now of age $j$ at time $t$.

The elements below the diagonal correspond to people who are infected with HIV and on ART so that the element $\mathbf{N}_i^j(t)$, $j < i$, gives the number of people who were infected with HIV at age $i$ and are on ART at age $j$.

Uninfected people may remain uninfected, they may die, or the may become infected. Infected people may remain infected, they may die, or they may start ART.

People on ART may remain on ART, may die or may fail treatment.

**Implementation**

To implement the programme we first set up the following vectors and matrices where all the elements are probabilities per unit time step.

$\mathbf{B}_i$: probability of dying, from causes unrelated to AIDS, for a person of age $i$;

$\mathbf{A}_i^j$: probability of dying from AIDS for a person infected at the age $i$ and now of age $j$.

$\mathbf{T}_i$: probability of starting treatment $i$ years after being infected.

$\mathbf{S}_i$: probability of failing (stopping) treatment $i$ years after being infected.

$\mathbf{F}_i$: the relative force of infection for a person of age $i$.

To initialize the (time-dependant) matrix $\mathbf{N}(t)$ we set $\mathbf{N}_i^i(0)$ to the age-distribution of people of age $i$ at time 0, normalized to 1. We set the crude birth rate $\beta$ and then choose the values of background mortality $\mathbf{B}_i$ so that the elements of $\mathbf{N}_i^i(0)$ match the current age-distribution assumed to be the stable age-distribution.

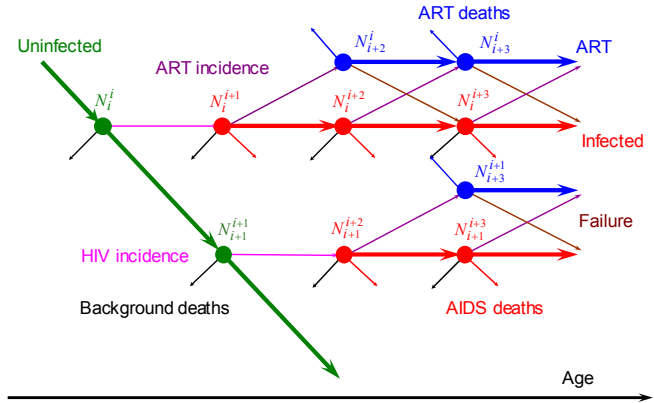

Figure 26. Labelling the states for reference to Equations 11 to 26. Heavy arrows are those that survive from one time to the next. Green: susceptible; red: infected; blue: treated. Light arrows are those that change their state. Black: background mortality; red: AIDS mortality; blue: mortality on treatment; pink: incident cases; purple: starting treatment; brown; stopping treatment.

We then define the following variables as a function of time $t$:

Number of susceptible people

$$S(t) = \sum_i \mathbf{N}_i^i(t) \qquad 11$$

Number of people infected with HIV, not on ART, (only those older than 15 can be infected)

$$H(t) = \sum_{i,j>i} \mathbf{N}_i^j(t) \qquad 12$$

Number of people on ART (treatment) (only those older than 15 can be infected)

$$T(t) = \sum_{i,j<i} \mathbf{N}_i^j(t) \qquad 13$$

Total population

$$N(t) = S(t) + H(t) + T(t) \qquad 14$$

Birth rate (children)

$$C(t) = \beta N(t) \qquad 15$$

Adult population

$$R(t) = \sum_{i,j>15} \mathbf{N}_i^j(t) \qquad 16$$

Adult prevalence of HIV

$$P(t) = H(t)/R(t) \qquad 17$$

Adult prevalence of ART

$$Q(t) = T(t)/R(t) \qquad 18$$

Incidence of HIV

$$I(t) = \lambda P(t) \sum_i \mathbf{F}_i \mathbf{N}_i^i(t) \qquad 19$$

Background mortality

$$BM(t) = \sum_{i,j} \mathbf{N}_i^j(t) \qquad 20$$

AIDS mortality

$$AM(t) = \sum_{i,j>i} \mathbf{A}_i^j \mathbf{N}_i^j(t) \qquad 21$$

**Iterating over time**

We first consider progression for those infected with HIV on and off ART (Figure 27).

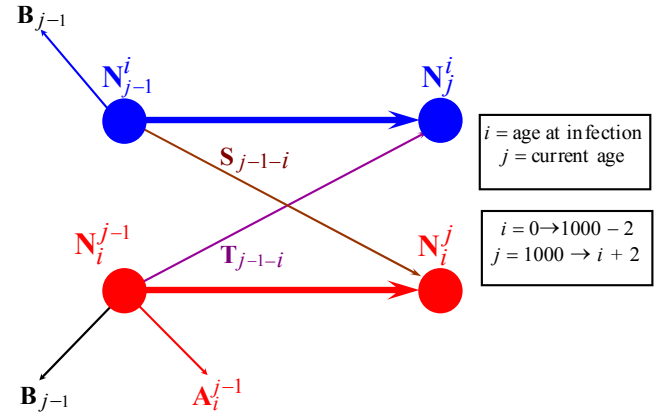

Figure 27. Infected people can die of natural causes (background mortality) (**B**), AIDS (**A**) or start treatment (**T**). People on treatment can die of natural causes (background mortality) (**B**) or stop treatment (**S**). Note that the red elements (infected with HIV) are stored above the diagonal; the blue elements (on ART) are stored below the diagonal. The diagonal elements are all susceptible and the elements immediately above the diagonal are incident cases. Both are dealt with in the text.

We now update the matrix of susceptible, infected and treated people. We first move from right to left along each pair of rows illustrated in Figure 27. At this stage we exclude the diagonal and the elements one above and one below the diagonal. If $d$ is the dimension of the matrix, the loop is

For $i = 0$ to $d - 2$ Step 1
    For $j = d$ to $i + 2$ Step $-1$

$$\mathbf{N}_i^j(t+1) = \mathbf{N}_i^{j-1}(t)\left[1 - \mathbf{B}^{j-1}(t) - \mathbf{A}_i^{j-1}(t) - \mathbf{T}_{j-1}(t)\right] + \mathbf{N}_{j-1}^i(t)\mathbf{S}_{j-1}(t) \qquad 22$$

$$\mathbf{N}_j^i(t+1) = \mathbf{N}_{j-1}^i(t)\left[1 - \mathbf{B}_{j-1}(t) - \mathbf{S}_{j-1}(t)\right] + \mathbf{N}_i^{j-1}(t)\mathbf{T}_{j-1}(t) \qquad 23$$

    Next
Next

This accounts for all transitions among people infected with HIV. We now have to deal with the incident infections.

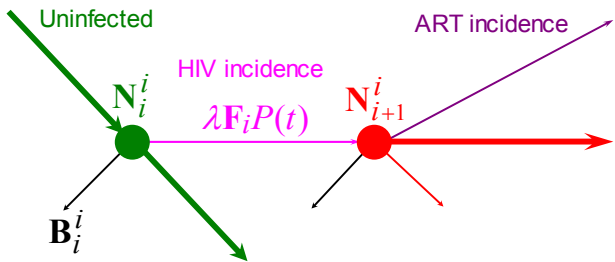

Figure 28. We have to allow for incident infections (pink arrow) and the aging of the susceptible population (green arrows)

We now update the incident cases:

For $i = 1$ to $d - 1$ Step 1

$$\mathbf{N}_{i+1}^i(t+1) = \lambda \mathbf{F}_i P(t) \mathbf{N}_i^i(t) \qquad 24$$

    Next

and then update the susceptible cases:

For $i = 1$ to $d - 1$ Step 1

$$\mathbf{N}_i^i(t+1) = \mathbf{N}_{i-1}^{i-1}(t)\left(1 - \lambda \mathbf{F}_i P(t) - \mathbf{B}_{i-1}^{i-1}\right) \qquad 25$$

    Next

and finally account for births:

$$\mathbf{N}_0^0(t+1) = \beta T(t) \qquad 26$$

## References

1. Einstein A. On the method of theoretical physics. *Philosophy of Science*. 1934; 1: 163-9.
2. Williams BG. Optimizing control of HIV in Kenya. arXiv 2014; Available from: http://arxiv.org/abs/1407.7801
3. Kenya Go. Kenya Open Data. 2014; Available at: https://www.opendata.go.ke.
4. Williams B, Gouws E, Wilkinson D, Abdool Karim S. Estimating HIV incidence rates from age prevalence data in epidemic situations. *Statistics in Medicine*. 2001; 20: 2003-16.
5. CASCADE Collaboration. Time from HIV-1 seroconversion to AIDS and death before widespread use of highly-active anti-retroviral therapy. A collaborative analysis. *Lancet*. 2000; 355: 1131-7.
6. Williams BG, Granich R, Chauhan LS, Dharmshaktu NS, Dye C. The impact of HIV/AIDS on the control of tuberculosis in India. *Proceedings of the National Academy of Sciences USA*. 2005; 102: 9619-24.
7. Williams BG, Gouws E. R0 and the elimination of HIV in Africa: Will 90-90-90 be sufficient? arXiv 2014; Available from: http://arxiv.org/abs/1304.3720
8. Garnett GP, Anderson RM. Sexually transmitted diseases and sexual behaviour: insights from mathematical models. *Journal of Infectious Diseases*. 1996; 174 Suppl 2: S150-61.
9. Granich RM, Gilks CF, Dye C, De Cock KM, Williams BG. Universal voluntary HIV testing with immediate antiretroviral therapy as a strategy for elimination of HIV transmission: a mathematical model. *Lancet*. 2008; 373: 48-57.
10. Mugyenyi P, Walker AS, Hakim J, Munderi P, Gibb DM, Kityo C, *et al*. Routine versus clinically driven laboratory monitoring of HIV antiretroviral therapy in Africa (DART): a randomised non-inferiority trial. *Lancet*. 2010; 375: 123-31.
11. Maman D, Pujades-Rodriguez M, Nicholas S, McGuire M, Szumilin E, Ecochard R, *et al*. Response to antiretroviral therapy: improved survival associated with CD4 above 500 cells/μl. *AIDS*. 2012; 26: 1393-8.
12. Hargrove JW, Humphrey JH, Mahomva A, Williams BG, Chidawanyika H, Mutasa K, *et al*. Declining HIV prevalence and incidence in perinatal women in Harare, Zimbabwe. *Epidemics*. 2011; 3: 88-94.
13. Williams BG, Gouws E. HIV and TB in Eastern and Southern Africa: Evidence for behaviour change and the impact of ART. arXiv 2014; Available from: http://arxiv.org/abs/1406.6912v1
14. Gill VS, Lima VD, Zhang W, Wynhoven B, Yip B, Hogg RS, *et al*. Improved virological outcomes in British Columbia concomitant with decreasing incidence of HIV type 1 drug resistance detection. *Clinical Infectious Diseases*. 2010; 50: 98-105.